\documentclass[twocolumn,trackchanges]{aastex63}

\usepackage{graphicx}	
\usepackage{float} 
\usepackage{amsmath}	
\usepackage{amssymb}	
\usepackage{natbib}

\shorttitle{Pixelation Effect in Shear Measurement}
\shortauthors{Shen et al.}

\graphicspath{{./}{figures/}} 
\begin{document} 
\title{Tolerance For the Pixelation Effect in Shear Measurement}
\correspondingauthor{Jun Zhang}
\email{betajzhang@sjtu.edu.cn}
\author[0000-0002-7825-6263]{Zhi Shen}

\affiliation{Department of Astronomy, Shanghai Jiao Tong University, Shanghai 200240, China}
\author[0000-0003-0002-630X]{Jun Zhang}
\affiliation{Department of Astronomy, Shanghai Jiao Tong University, Shanghai 200240, China}
\affiliation{Shanghai Key Laboratory for Particle Physics and Cosmology, Shanghai 200240, China}
\author[0000-0002-0610-2361]{Hekun Li}
\affiliation{Department of Astronomy, Shanghai Jiao Tong University, Shanghai 200240, China}
\author[0000-0001-5945-8399]{Haoran Wang}
\affiliation{Department of Astronomy, Shanghai Jiao Tong University, Shanghai 200240, China}
\author[0000-0001-5912-7522]{Chengliang Wei}
\affiliation{Purple Mountain Observatory, Chinese Academy of Sciences, Nanjing, 210023, China}
\author{Guoliang Li}
\affiliation{Purple Mountain Observatory, Chinese Academy of Sciences, Nanjing, 210023, China}
\affiliation{School of Astronomy and Space Science, University of Science and Technology of China, Hefei, 230026, China}
\author{Xiaobo Li}
\affiliation{Changchun Institute of Optics, Fine Mechanics and Physics, Chinese Academy of Sciences, Changchun, 130033, China}
\author{Zhang Ban}
\affiliation{Changchun Institute of Optics, Fine Mechanics and Physics, Chinese Academy of Sciences, Changchun, 130033, China}
\author{Dan Yue}
\affiliation{College of Physics, Changchun University of Science and Technology, Changchun 130022, China;}
\begin{abstract}
Images taken by space telescopes typically have a superb spatial resolution, but a relatively poor sampling rate due to the finite CCD pixel size. Beyond the Nyquist limit, it becomes uncertain how much the pixelation effect may affect the accuracy of galaxy shape measurement. It is timely to study this issue given that a number of space-based large-scale weak lensing surveys are planned. Using the Fourier\_Quad method, we quantify the shear recovery error as a function of the sampling factor Q, i.e., the ratio between the FWHM of the point-spread-function (PSF) and the pixel size of the CCD, for different PSFs and galaxies of different sizes and noise levels. We show that sub-percent-level accuracy in shear recovery is achievable with single-exposure images for $Q\lesssim 2$. The conclusion holds for galaxies much smaller than the PSF, and those with a significant level of noise.
\end{abstract} 



\section{Introduction}\label{sec:intr}

Weak gravitational lensing is now widely known as a powerful probe of the cosmic large scale structure \citep{Hoekstra2008,Kilbinger2015,Mandelbaum2018}. 
For the purpose of placing tighter constraints on the cosmological parameters, a number of large scale galaxy surveys in space are planned, including Euclid \citep{Laureijs2011}, the China Space Station Telescope (CSST) \citep{Gong2019}, and Roman \citep{roman}, all of which are going to observe billions of galaxy images for accurate weak lensing measurement. 

\begin{figure}
	\centering
	\includegraphics[scale = 0.7]{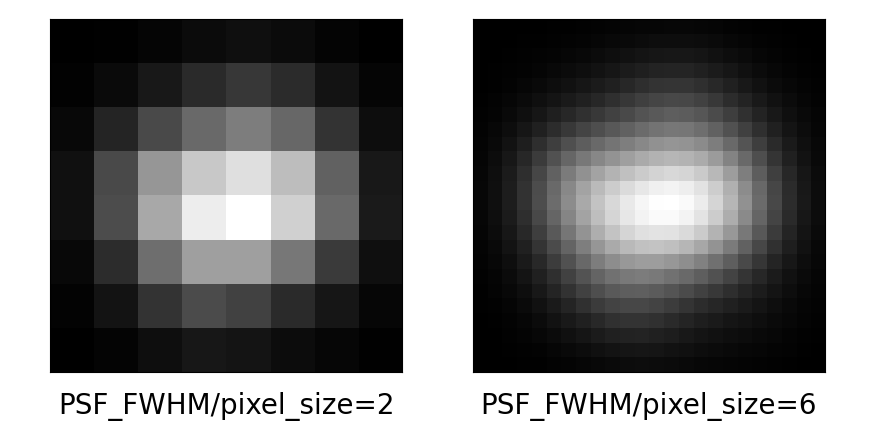}
	\caption{Undersampled (left) and oversampled (right) images for the same galaxy.}
	\label{fig:uwimgs}
\end{figure}

The main advantage of space telescope is its superb spatial resolution due to the lack of atmospheric turbulence and low sky background. One can therefore resolve more distant or fainter galaxies, and enhance the statistical power in weak lensing studies. In doing so, one particular challenge is to deal with the so-called pixelation effect, i.e., the CCD pixel size is not small enough with respect to the size of the point spread function (PSF). For example, in Fig.\ref{fig:uwimgs}, we show the images of the same mock galaxy sampled with two different pixel sizes. It is interesting and important to ask to what extent would shear measurement tolerate the discreteness of the images. 

The pixelation effect has been specifically discussed for several shear measurement methods. For example, \cite{High2007} uses RRG, a modification of the KSB+ \citep{Kaiser1995,Hoekstra1998}, to measure the moments of the \textit{Drizzled} images made of coadded undersampled images \citep{Fruchter_2002}. \cite{Zhang2015} studies the performance of the Fourier\_Quad shear measurement method (FQ hereafter) with the coadded COSMOS images that are processed also with the Drizzle algorithm. More general ways of linearly combining images have been proposed and tested by \cite{Rowe_2011} and \cite{Shapiro_2013}. For combined images, the success of shear measurement relies on several important conditions, including (but not limited to) the following: the homogeneity of pixel alignment, the temporal and spatial variation of the PSF, the pointing accuracy of each exposure on subpixel scales, the difference in the optical distortion. A comprehensive study of these issues is very complicated, and sometimes requires specific knowledge of the survey that is often inaccessible for general users of the data. 

More recently, \cite{Kannawadi2021} studies the performance of KSB \citep{Kaiser1995} and re-Gaussianization \citep{HirataSeljak2003} combined with the metacalibration \citep{Huff-metacal,Sheldon-metacal} algorithm using images simulated for the EUCLID survey. The metacalibrated exposures are combined to form oversampled images for shear measurement using either KSB or re-Gaussianization. It is a promising direction, but we are aware that the inter-pixel interpolation is very time consuming ($\gtrsim$ 1CPU*sec/galaxy), and the bias could be large for sources of low signal-to-noise ratio ($\lesssim 20-30$).

In this paper, we take a simpler and more straightforward approach: we study the performance of FQ for images on individual exposures, without involving any image coadding process. This topic is indeed highly relevant to CSST, which is a major science project established by the space application system of the China Manned Space Program \citep{Zhan2011,Cao2018}. CSST is a 2 m space telescope in the same orbit as the China Manned Space Station, and is planned to be launched at the end of 2023. The CSST weak lensing survey will cover about 17,500 deg$^2$ sky area with survey depth reaching i$\approx 26$ AB magnitude (5$\sigma$ detection for point sources). It contains seven filters: NUV, u, g, r, i, z, and y bands, covering the wavelength range 255–1000 nm with high spatial resolution ($\sim 0.15$ arcsec, radius of 80\% energy concentration region). The current plan of the CSST main survey is to take only two exposures per object in each band on two separate CCDs. For this observing strategy, we believe it is very challenging to perform accurate shear measurement with coadded images (although multiple exposures in either the same or different bands are still very useful in, e.g., cosmic ray identifications). 

On single-exposure images, it has been previously demonstrated with theoretical reasonings and numerical simulations \citep{Zhang2008,Zhang2010,zhang2011JCAP,Zhang2011,Zhang2015,Zhang2017PDF} that the FQ method is quite robust in many different aspects of shear measurement, including: 

1. It does not contain any assumptions about the morphologies of the galaxies or the PSF's. Because of this fact, in principle, the method does not require calibration in the absence of detector effects.

2. The image processing steps are simple and straightforward, not involving pixel-level manipulations such as interpolation.

3. The method includes rigorous treatment of noise, including the background and the Poisson noise.

4. Numerically, the calculations in FQ require mainly the Fast Fourier Transformation, which only takes about $10^{-3}$ CPU*sec/galaxy, much faster than, e.g., typical model-fitting methods.  

The accuracy of FQ has also been tested with the CFHTLenS and DECaLS data in \cite{Zhang_2019} and \cite{Wang_2021}, in which we find that the galaxy shears are in good agreement with the small field distortion signals ($0.1\%-0.5\%$).

In \S\ref{sec:2}, we give a brief introduction of the FQ method, and show the performance of FQ as a function of the pixel size under different assumptions about the PSF form, the galaxy size, and the noise level. In \S\ref{remedy}, we discuss a simple way of reducing the impact of the pixel size on shear measurement. Finally, we summarize our findings in \S\ref{summary}.


\section{Pixelation effect on shear measurement}\label{sec:2}
\subsection{Shear measurement method}\label{sec:2.1} 
   
In this work, we focus on the impact of the pixelation effect on the Fourier\_Quad shear measurement method. The FQ method utilizes the multipole moments of the 2D galaxy power spectrum to recover the cosmic shear signal. Its shear estimators are defined as:
        \begin{equation}
            \begin{aligned}
                G_1&=-\frac{1}{2}\int d^2 \vec{k}(k_x^2-k_y^2)T(\vec{k})M(\vec{k})\\  
                G_2&=-\int d^2\vec{k}k_x k_y T(\vec{k})M(\vec{k})\\
                N&=\int d^2 \vec{k}\left[k^2-\frac{\beta^2}{2}k^4\right]T(\vec{k})M(\vec{k})
            \end{aligned}
        \end{equation}
$M(\vec{k})$ is the 2D galaxy power spectrum corrected by terms related to the background noise and the Poisson noise (see eq.(4.9) of \cite{Zhang2015}). $T(\vec{k})$ is the factor for converting the PSF to a Gaussian form, i.e.:   
\begin{equation}
T(\vec{k})={\big|\widetilde{W}_{\beta}(\vec{k})\big|}^2/{\big|\widetilde{W}_{PSF}(\vec{k})\big|}^2\label{tk}
\end{equation}
in which $\widetilde{W}_{PSF}(\vec{k})$ and $\widetilde{W}_{\beta}(\vec{k})$ $[=\exp(-\beta^2\big|\vec{k}\big|^2/2)]$ are the Fourier transforms of the PSF function and the Gaussian kernel respectively. The value of $\beta$ should be chosen to be somewhat larger than the scale radius of the PSF to avoid numerical instability in the conversion. In most of the examples shown in this paper, the value of $\beta$ is chosen so that the FWHM of the Gaussian kernel is 1.4 times that of the original PSF. Later in this article, we show that the choice of $\beta$ can also help to reduce the pixelation effect in shear measurement. Using the shear estimators defined above, one can show that the shear signal can be recovered to the second order in accuracy \citep{Zhang2015}, i.e.:
\begin{equation}
\label{ave}
\frac{\langle G_i\rangle}{\langle N\rangle}=g_i+O(g^3_{1,2})\quad \quad (i=1,2).
\end{equation}
Moreover, other than taking the ensemble averages of the shear estimators, as what is conventionally done, one can in principle recover the shear more accurately by utilizing the full information in the probability distribution function of the estimators, as shown in \cite{Zhang_2016}. Nevertheless, we simply use the ensemble averaging method in eq.\ref{ave} to derive all the results in this paper, as we find that all of our conclusions are not affected by the choice of the statistical method.

\subsection{General setup for image simulations}\label{sec:2.2}
 
We set up two types of simulations: 1. regular galaxies with \textsl{De Vaucouleurs} type profile generated by \textit{GalSim} \citep{Rowe2015}; 2. irregular galaxies made of point sources whose positions are determined by 2D random walks \citep{Zhang2008}. In most of our discussions below, we use the irregular galaxies for several reasons: 1. they are very fast and convenient to generate; 2. their irregular shapes make our conclusion more robust; 3. numerical operations on these images (shearing, convolving with PSF) only involve point sources, therefore they can be done very accurately.
 
 
 
\begin{figure}
	\centering
	\includegraphics[scale =1]{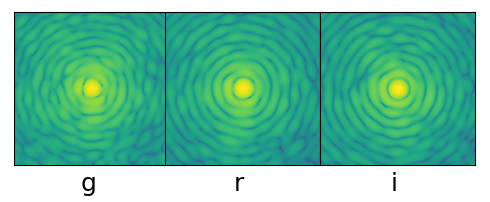}
	\caption{Simulated PSF profiles for the g/r/i bands of CSST (in logarithmic scale).}
	\label{fig:csstpsf}
\end{figure}

We consider four types of PSF: Gaussian, Moffat, Airy function, and the mock PSF (in g/r/i three bands) generated with realistic optics of CSST. 
The PSF images of CSST are shown in Figure.\ref{fig:csstpsf}.
In order to obtain a set of realistic PSFs to account for the impact of optical system on image quality, the CSST image simulation team has developed an optical emulator to produce high fidelity mocked PSF of CSST.
The optical emulator of CSST is based on six different modules to simulate the optical aberration due to mirror surface roughness, CCD assembly errors, fabrication errors, gravitational and thermal distortions. Moreover, two dynamical errors, due to micro-vibrations and image stabilization, are also included in the simulated PSF. 

Finally, since the simulations in this work often have a large pixel size with respect to the PSF, we need to take into account the so-called pixel response (PRF hereafter)\citep{High2007}. For convenience, we set the PRF to be a normalized square tophat $R_{sq}(\theta)$ :
        \begin{equation}
            R_{sq}(\theta)=
            \left\{
                 \begin{array}{lr}
                 \theta^{-2}_{CCD} & if|\theta_x|,|\theta_y|<\theta_{CCD}/2,  \\
                 0                 & otherwise
                 \end{array}
            \right.
        \end{equation}
The PRF is convolved with the original PSF to form the effective PSF. In the rest of this paper, the PSF we use always refers to the effective PSF. \\
        
        
\subsection{A simple test}\label{test}

A useful quantity for characterizing the level of the pixelation effect is the ratio between the full-width-at-half-maximum (FWHM) of the PSF and the pixel size, which we define as the sampling factor Q. According to the sampling theorem, $Q=2$ is a critical point, below which the sampling rate is deemed not sufficient. For this reason, most of our results in this work are presented as a function of Q.

\begin{figure}
  \centering \includegraphics[scale=0.5]{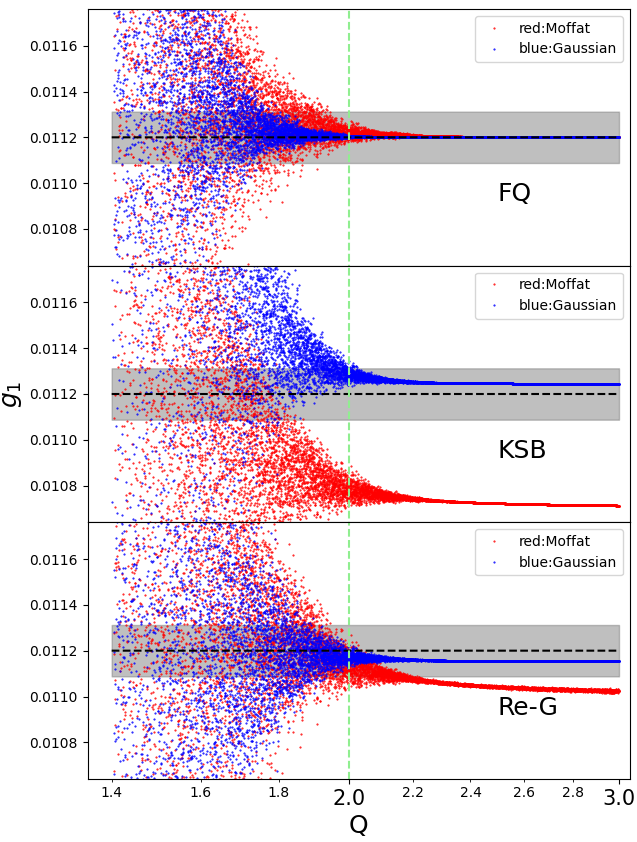}
  \caption{Performances of the three shear measurement methods (FQ, KSB, re-Gaussianization) for different values of Q. The dashed lines indicate the input shear values, and the shaded region represents the upper and lower 1\% bound of input shear values. Each data point shows the recovered shear value using four images (separated by 45 degrees to remove the shape noise and the anisotropic shear responsivity) of one mock galaxy generated by \textit{Galsim}. No noise is added. In each panel, we include the results for the Gaussian and Moffat PSF in blue and red respectively.}
  \label{fig:galsim}
\end{figure}

As our first example, we show briefly how the performance of shear measurement changes when the pixel size becomes increasingly large comparing to the PSF size. For simplicity, we only use the Gaussian and Moffat functions for the PSF in this example. In Fig.\ref{fig:galsim}, we show the results from three shear measurement methods: FQ, KSB, and re-Gaussianization. KSB is a method based on the weighted multipole moments of the galaxy image in real space, with the capability of correcting for the PSF effect \citep{Kaiser2000}. re-Gaussianization is another method which uses the perturbative Gaussian deviations to correct for the non-Gaussianity of both PSF and galaxies \citep{Hirata2003,Mandelbaum2006a}. We use the implementation of KSB and re-Gaussianization in \textit{GalSim} to conduct the test. In this test, the galaxy images are generated by \textit{GalSim} as well. For the results in each panel, we generate $10^4$ galaxies (without noise). Each data point in the figure stands for the average shear result from four images of a single galaxy separated by 45 degrees in rotation, for the purpose of suppressing the shape noise as well as the anisotropic part of the shear response (spin-4 component) \citep{Zhang2011}. For each galaxy, its PSF size is randomly chosen so that the Q value is in the range of $[1.4, 3]$. 

For simplicity, we only show the results of $\hat{g}_1$. For the results of KSB and re-Gaussianization, no corrections from shear responsivities are included, as they are not important for our purpose here. Fig.\ref{fig:galsim} shows that the recovered shear signal starts to diverge when $Q\lesssim 2$ for all three methods. These results are in agreement with the sampling theorem. For either CSST or EUCLID, we expect the Q value to be less than or close to 2. It is therefore critical to understand the precision of shear recovery at the neighborhood of $Q=2$ for these projects.

\subsection{Performance of FQ on different PSFs}\label{sec:3}
        
\begin{figure}
	\centering
	\includegraphics[scale = 0.37]{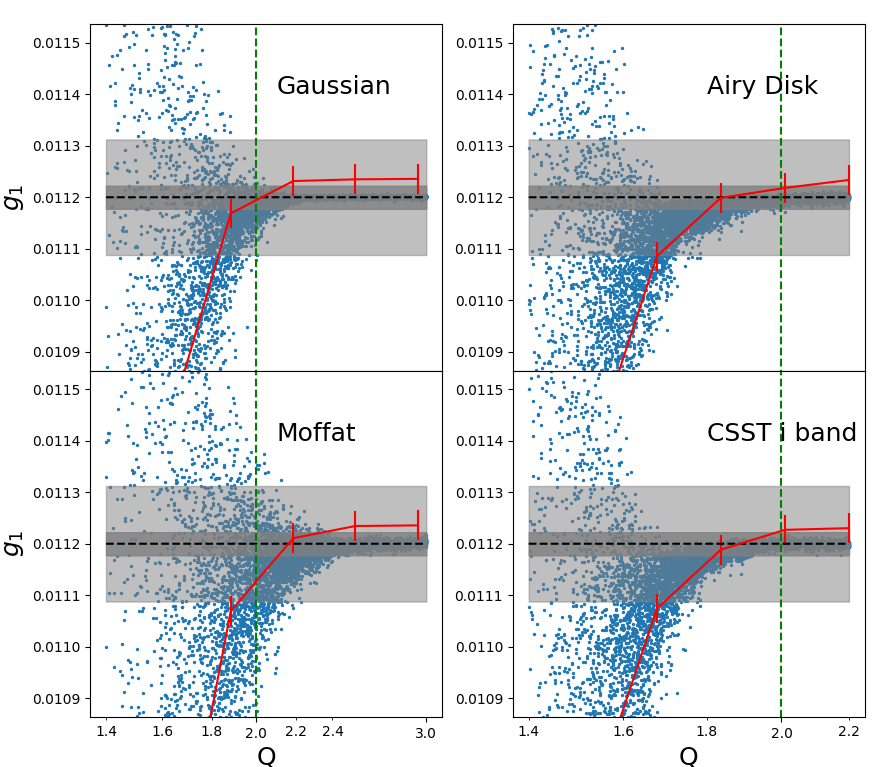}
	\caption{Performance of FQ for four types of PSFs: Gaussian, Moffat, Airy disk, CSST i-band. The blue data points and light grey regions are similarly defined as in Fig.\ref{fig:galsim}. The dark grey regions show the $2*10^{-3}$ bound around the input shear value. The red data points with error bars are the ensemble averages of the shear value, which indicate the level of systematic bias in the presence of the pixelation effect. The galaxies used in producing this figure are all generated using random walks of point sources.}
	\label{fig:gmai10000}
\end{figure}

Let us now only focus on the FQ method. Similar to Fig.\ref{fig:galsim}, we show the shear recovery results for the FQ method with two additional PSF types in Fig.\ref{fig:gmai10000}. In these tests, we use irregular galaxies. In addition to the scattered data points, in each panel we use an ensemble of $10^7$  galaxies (without rotations for the cancellation of the shape noise) to produce the ensemble average of the shear value as a function of Q, shown as the red data points with error bars in the plot.  We can see that the systematic errors shot up slightly later than the statistical errors when Q decreases. An important finding is that FQ seems to have more tolerance on the pixelization effect when the PSF is Airy disk or something alike (CSST's PSF). 

\begin{figure} 
	\centering  \includegraphics[scale=0.35]{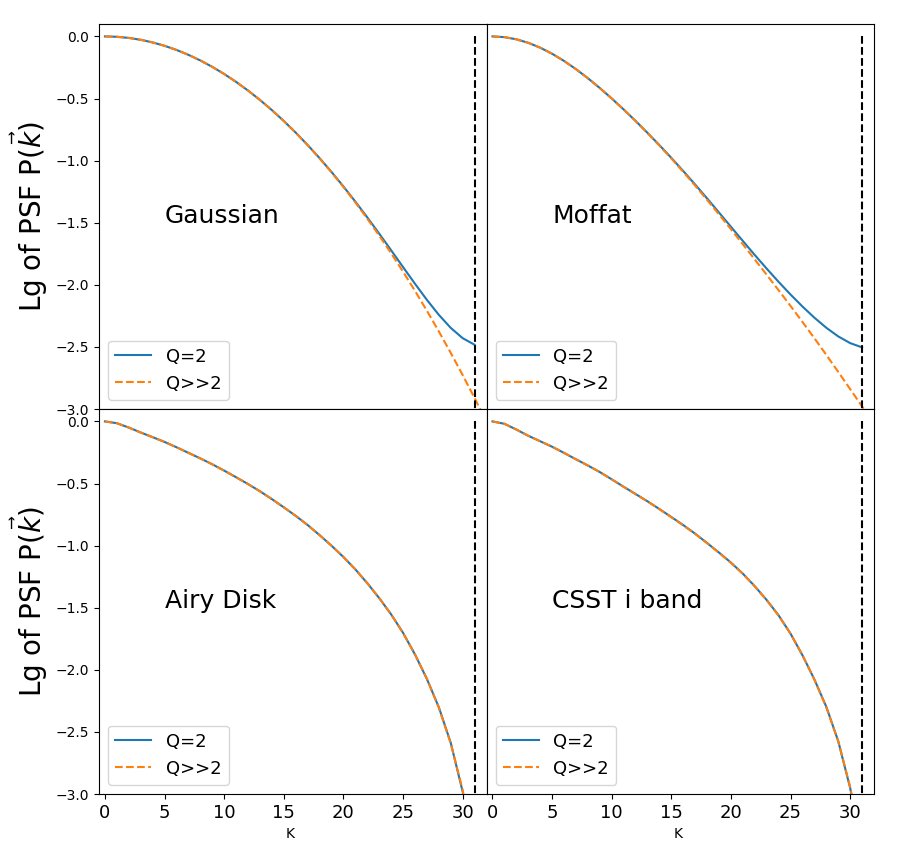}
	\caption{The blue line in each panel shows the PSF power spectrum for $Q=2$, and the orange dashed line is the PSF power for $Q>>2$, i.e., without the aliasing power.}
	\label{fig:gma-als-p}
\end{figure}

To understand the difference between the Airy disk PSF and the Gaussian or Moffat functions, we plot the influence of the aliasing power in all cases in Fig.\ref{fig:gma-als-p}. The blue solid lines in the figure are the power spectra of the PSFs for Q = 2, and the orange dashed lines are what the power would be if there were no aliasing issues (calculated by choosing a smaller pixel size, and then rescale the wave number). It can be found that Airy function is much less affected by the aliasing power than Gaussian or Moffat when Q=2. This is perhaps not surprising: the Airy disk is the square of the Fourier transform of a circular aperture, therefore its Fourier transform is the convolution of two circular aperture, which is quite localized in the Fourier domain comparing to that of either the Gaussian or the Moffat function. This is an encouraging news, as we know that for space-based observations, the PSF is close to the form of an Airy disk. 

\subsection{Quantifying the systematic error}\label{sec:3.2}


In this section, we study the other factors that can affect the pixelation effect, including the galaxy size and the signal-to-noise ratio (SNR). In order to calibrate the systematic error in shear recovery, we adopt the commonly used multiplicative bias m and additive bias c \citep{Heymans2006} defined as:
\begin{equation}
\hat{g}_i=(1+m_i)g^{true}_i+c_i, (i=1,2)
\end{equation}
where $\hat{g}_i$ and $g^{true}_i$ are the recovered and true shear values respectively. To measure m and c, we use 10 sets of random input shear values in the range of [-0.02, 0.02]. 

\begin{figure}
	\centering
	\includegraphics[scale=0.5]{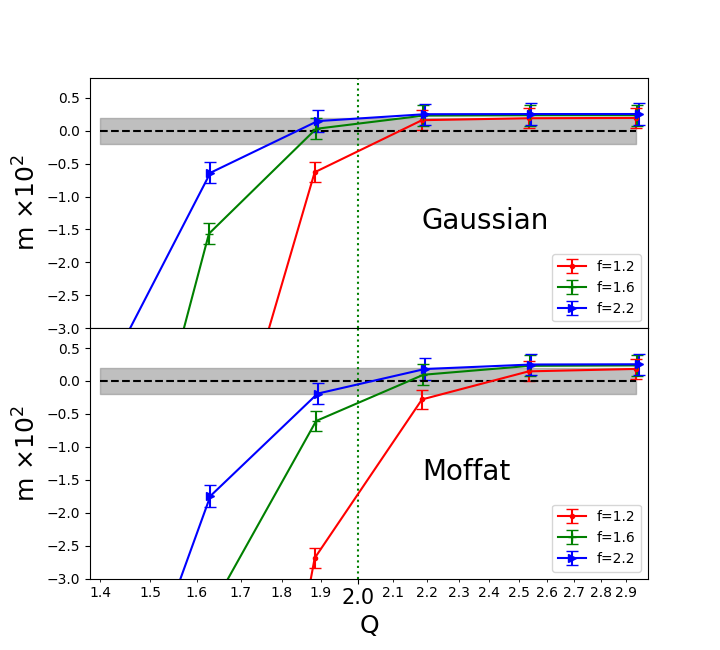}
	\caption{The multiplicative bias as a function of Q for different galaxy sizes. The two panels show the results for the Gaussian and Moffat PSF respectively. In each panel, the red, green, and blue lines are from galaxies of small, middle, and large sizes respectively. The parameter f is the ratio between the post-seeing FWHM of the galaxy and that of the PSF. Due to the similarity of $m_1$ and $m_2$, we just plot the average of $m_1$ and $m_2$ in this plot. The shaded region represents the requirement on m ($\pm 2*10^{-3}$) by the Stage IV surveys.}
	\label{fig:gma-m-size}
\end{figure}

\begin{figure}
	\centering
	\includegraphics[scale = 0.3]{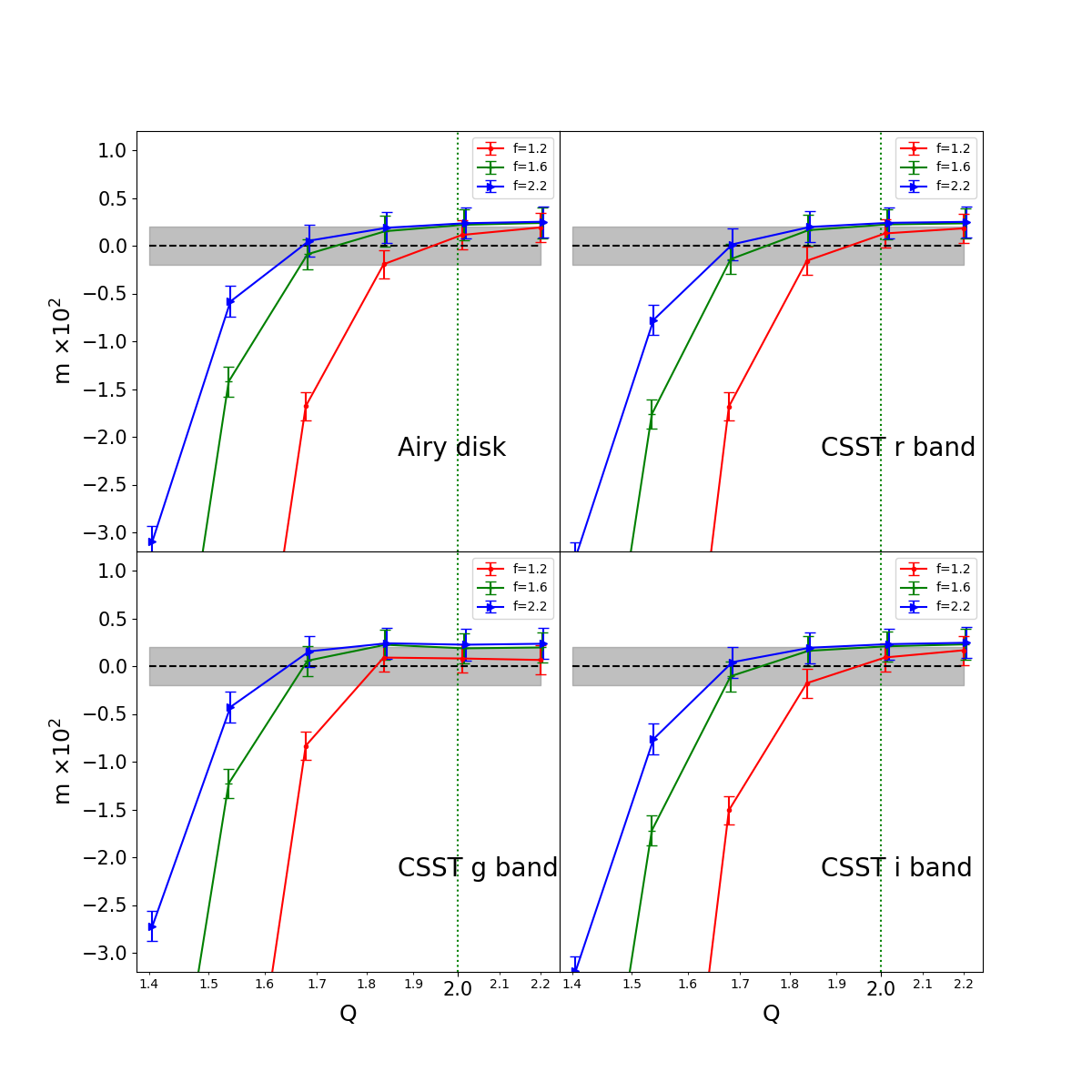}
	\caption{Similar to Figure.\ref{fig:gma-m-size}, but for the PSF of Airy disk and CSST.}
	\label{fig:csst-m-size}
\end{figure}

In Fig.\ref{fig:gma-m-size} and Fig.\ref{fig:csst-m-size}, we show the multiplicative biases from noise-free galaxies of different sizes. The results are shown for three different values of \textit{f}, which is defined as the ratio between the FWHM of the galaxy profile (PSF convolved) and that of the PSF. Each data point in these plots is measured using $10^8$ random-walk galaxies. To make the comparison easy, we choose the same random seed for the results of a given f in each panel of the figures. The shaded regions in these figures represent the range of m being within $\pm 2*10^{-3}$, a requirement by the Stage IV surveys \citep{Mandelbaum2018}. We adopt the same convention in the rest of the paper. In these tests, the additive bias c is always consistent with zero (due to the isotropy of the PSF), therefore we do not show it here. The figures show that for galaxies of larger sizes, the shear bias is somewhat smaller for a given value of Q. This is expected, as large galaxies covers more pixels each than smaller ones, their shape information on large scales should be less susceptible to the pixel size. However, even in the worst case (f=1.2), we find that FQ is accurate enough for the Airy disk and CSST PSF at $Q\approx 2$, which are most relevant to the upcoming space-based surveys. Note that for f=1.2, the pre-seeing galaxy size is already much smaller than the PSF size. 

\begin{figure}
     	\centering
     	\includegraphics[scale =0.5]{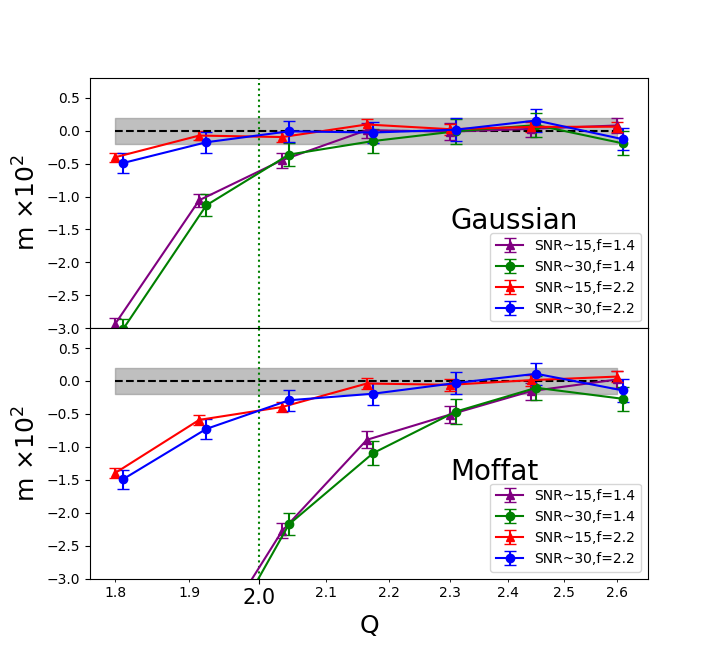}
     	\caption{The multiplicative bias for galaxies of different signal-to-noise ratios (SNR) and sizes. The two panels show the results for the Gaussian and Moffat PSF respectively. }
     	\label{fig:gma-m-snr}
\end{figure}

Furthermore, we consider how noise influences the pixelation effect.  We set two subsets with two choices of SNR: 15 and 30 (Poisson noise). The results for the Gaussian and Moffat PSFs are shown in Fig.\ref{fig:gma-m-snr}. For the case of SNR=15 and 30, we use $4*10^8$ and $10^8$ random-walk galaxies respectively to generate each data point. It is very encouraging to note that SNR does not strongly affect the performance of FQ in the presence of the pixelation effect.    
     
\begin{figure}
	\centering
	\includegraphics[scale = 0.5]{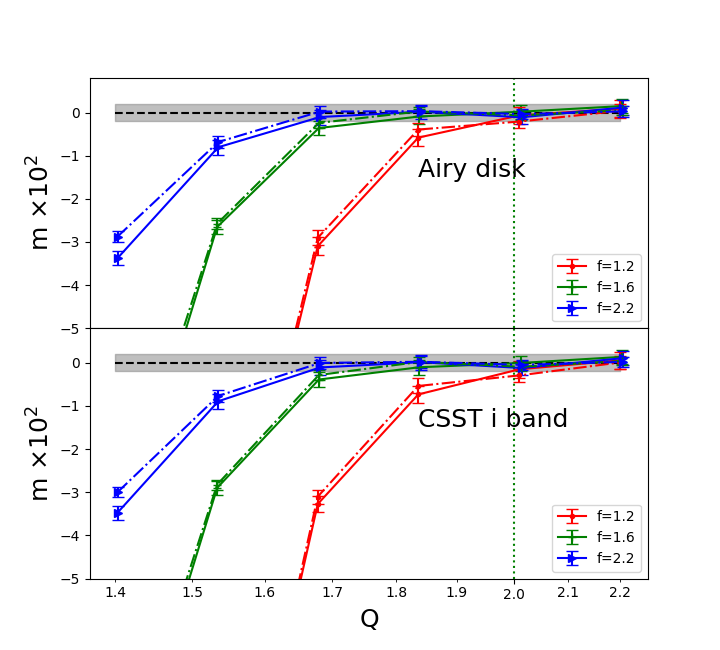}
	\caption{The multiplicative bias for galaxies of different sizes and noise level. The solid lines and the dash-dotted lines are from galaxies of SNR=30 or 15. The two panels are for the PSF of elliptical Airy disk and CSST respectively.}
	\label{fig:ai-m-e-size}
\end{figure}

\begin{figure}
	\centering
	\includegraphics[scale =  0.5]{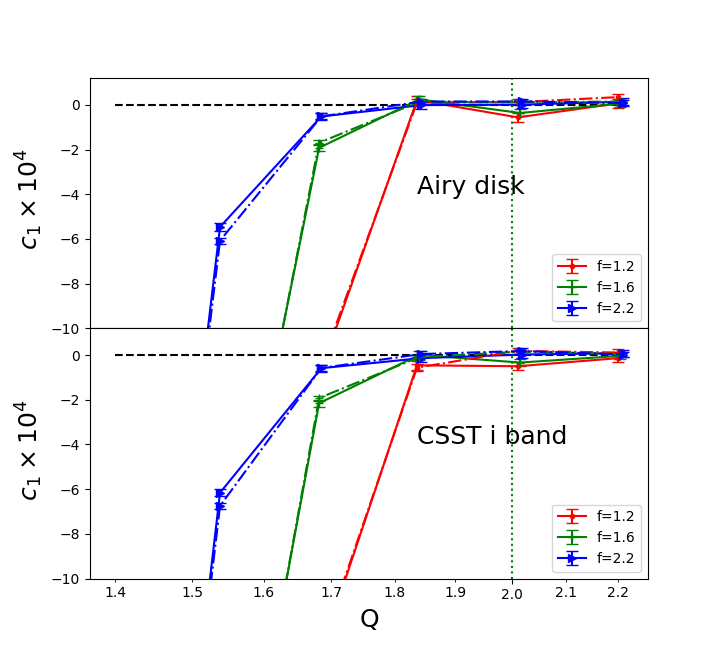}
	\caption{Similar to Figure.\ref{fig:ai-m-e-size}, but for the additive bias $c_1$.}
	\label{fig:ai-c-e-size}
\end{figure}

More generally, we consider using anisotropic PSFs in the test. In this case, the shear bias contains an additive part besides the multiplicative one. We show the Q-dependence of these two types of biases in Fig.\ref{fig:ai-m-e-size} and \ref{fig:ai-c-e-size} for the CSST i-band PSF and Airy disk. All the PSFs have 10\% ellipticity. The results in these two figures are from simulated galaxies with SNR=15 (solid lines) and 30 (dashed lines). The different colors refer to different galaxy sizes. One can see that the additive bias has a similar trend as that of the multiplicative bias when $Q$ decreases. Again, the transition point is not quite affected by the noise level. These findings demonstrate the robustness of FQ in realizing sub-percent-level accuracy in shear recovery for surveys like CSST with $Q\approx 2$ or even somewhat smaller. 

\section {A Remedy For the Pixelation Effect} 
\label{remedy}

We have demonstrated quite generally that the shear bias rises when the sampling factor $Q$ drops below a certain value, i.e., when there is a significant information loss on the sub-pixel scale. A possible way of remedying this problem is to re-convolve the galaxy with a larger kernel, so that the effective PSF size becomes larger, and the small scale information (which has been lost) would play a much less important role in shape measurement. Indeed, this idea can be straightforwardly implemented in FQ by changing the value of $\beta$ in eq.\ref{tk}. We are aware that a similar idea has been discussed in \cite{Kannawadi2021}.

\begin{figure}
	\centering
	\includegraphics[scale=0.5]{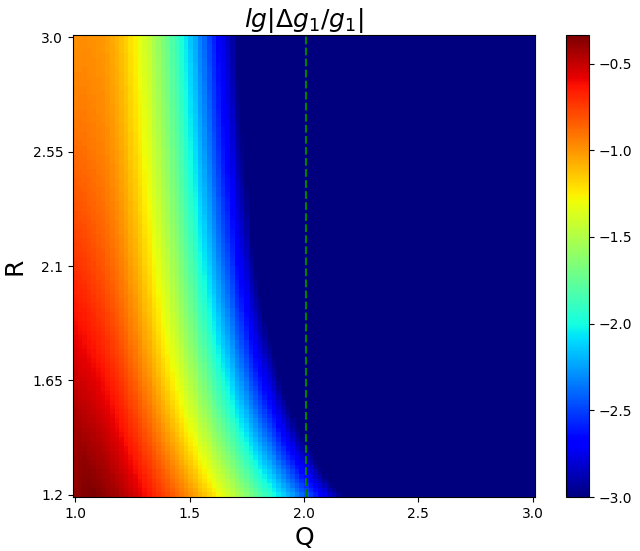}
	\caption{The relative error of the shear component $g_1$ as a function of Q and R. The PSF is set to be Gaussian. }
	\label{fig:gau_diag}
\end{figure}

\begin{figure}
	\centering
	\includegraphics[scale=0.5]{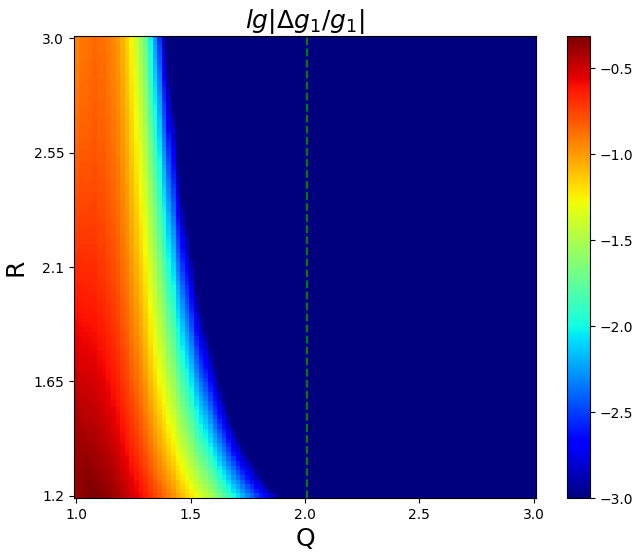}
	\caption{Similar to Figure.\ref{fig:gau_diag}, but for the PSF of Airy disk.}
	\label{fig:airy_diag}
\end{figure}

To find out how, let us define $R$ as the ratio between the FWHM of the Gaussian kernel and that of the original PSF in FQ. In Fig.\ref{fig:gau_diag} and \ref{fig:airy_diag}, we show how the shear recovery bias is affected by the choice of $R$ and $Q$ for the Gaussian and Airy disk PSF respectively. Every data point in the figures is estimated from 100 random-walk galaxies (f=2.2), for each of which we generate four copies of its image separated by 45 degrees to suppress the shape noise. In both figures, one can see clearly that for $Q$ being in the neighborhood of $2$, the shear bias can be significantly reduced by increasing $R$, as we expect. 

\begin{figure}
	\centering
	\includegraphics[scale = 0.35]{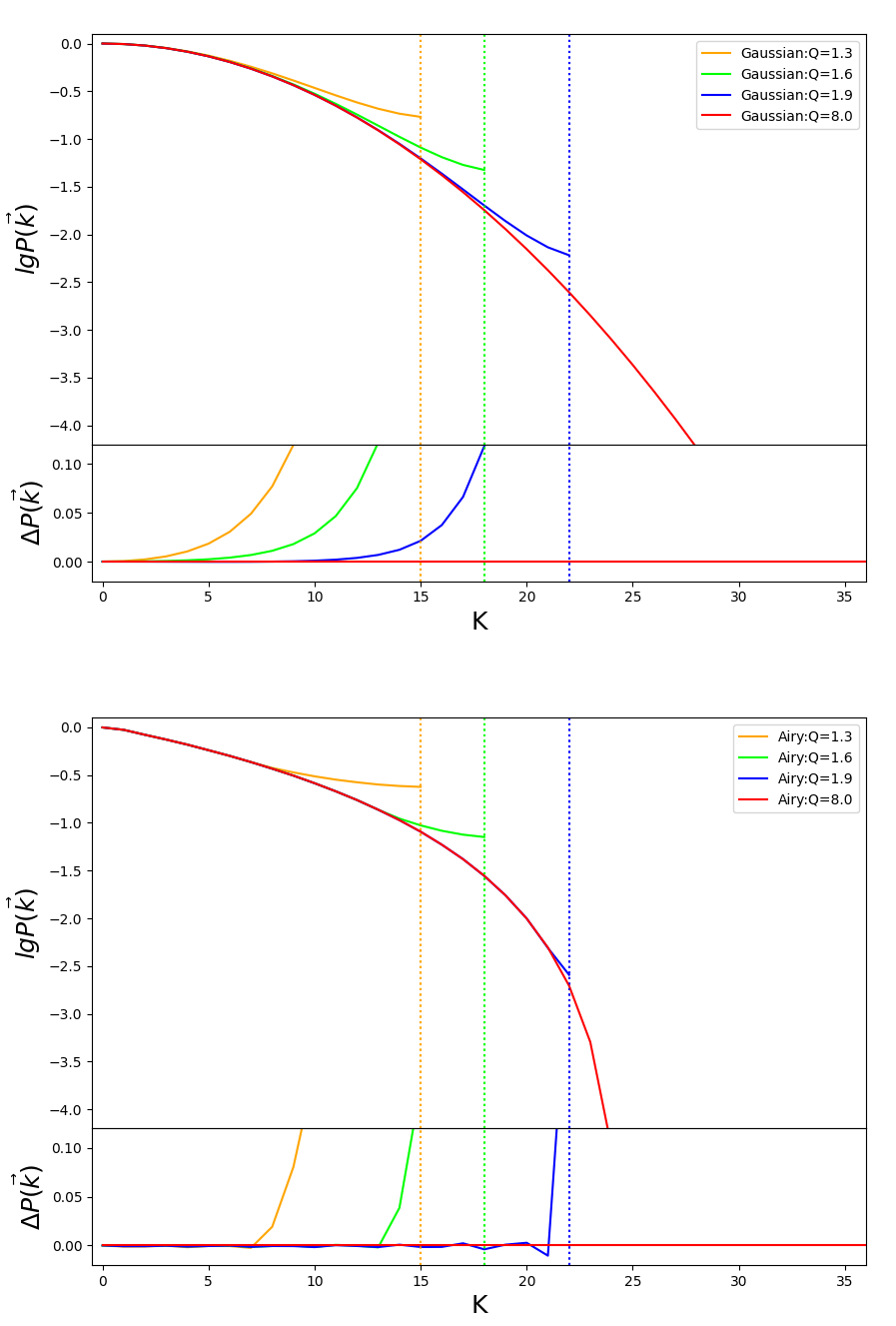}
	\caption{The level of aliasing power at different values of Q. The upper and lower panels show the results for the Gaussian and Airy disk PSF respectively. In each panel, the red line can be regarded as the power spectrum free of aliasing power, therefore treated as a reference.}
	\label{fig:gaPdps}
\end{figure}
In the FQ method, a larger value of $R$ means using a smaller central region of the Fourier space to infer the galaxy shape. The central region of the Fourier space is least affected by the aliasing power, as shown in Fig.\ref{fig:gaPdps}. We therefore can reduce the impact of pixelation by increasing $R$. However, this operation becomes much less useful when $Q$ reaches some critical value ($\approx 1.8$ for Gaussian and $\approx 1.3$ for Airy if f=2.2), i.e., when the innermost region of the Fourier space is contaminated by the aliasing power. For a given $Q$, compared to the Airy disk, the Gaussian PSF has a more extended area impacted by the aliasing power in Fourier space, making it more susceptible to the pixelation effect. In contrast, the Airy disk has remarkable compactness in Fourier space, therefore more restrained contamination from aliasing. This is consistent with what we found earlier in the paper.

\begin{figure}
	\centering
	\includegraphics[scale = 0.5]{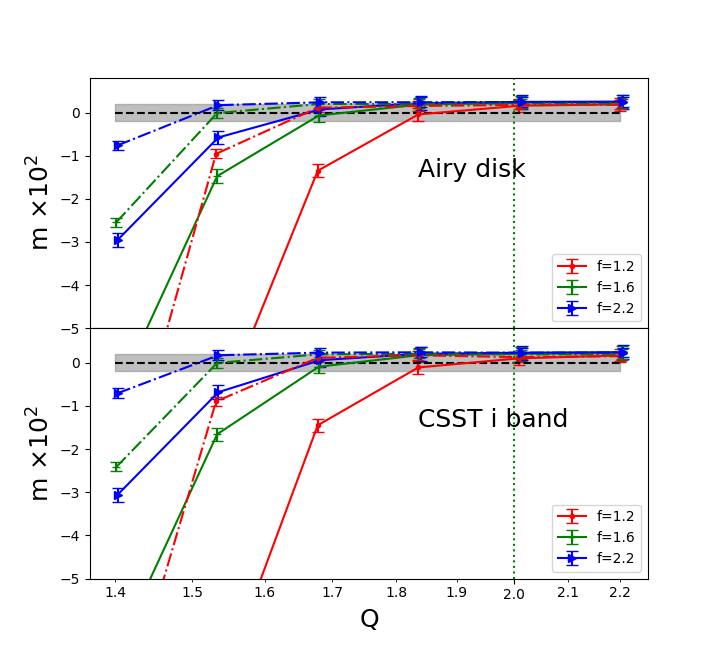}
	\caption{The multiplicative bias for galaxies of different sizes and R. The solid lines are from R=1.4, and the dash-dotted lines are from R=2.0. The two panels are for the PSF of Airy disk and CSST respectively.}
	\label{fig:ai-m-e-size-R}
\end{figure}

\begin{figure}
	\centering
	\includegraphics[scale =  0.5]{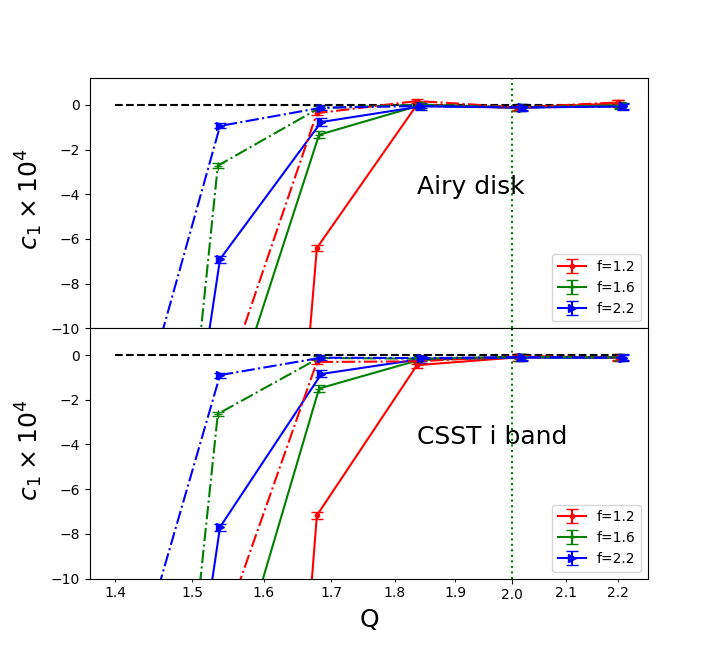}
	\caption{Similar to Figure.\ref{fig:ai-m-e-size-R}, but for the additive bias $c_1$.}
	\label{fig:ai-c-e-size-R}
\end{figure}

In Fig.\ref{fig:ai-m-e-size-R} and Fig.\ref{fig:ai-c-e-size-R} we show how multiplicative and additive bias are affected by $R$ for galaxies of different sizes. In these results, the galaxies are free of noise. The solid and dashed lines are the results for $R$ being 1.4 and 2.0 respectively. It is clear that by increasing $R$, one can significantly increase the accuracy of shear recovery. Although, we note that the value of $R$ shall be limited by the stamp size of the source.  


\section{Conclusion}\label{summary} 

In space-based observations, the pixelation effect can lead to significant shear measurement bias when the sampling factor is too low ($Q\lesssim 2$). Using three methods (FQ, KSB and re-Gaussianization), we have shown in Fig.\ref{fig:galsim} that in general, certain instability arises in galaxy shape measurement when the pixel size is too large comparing to the PSF FWHM. We then specifically study the Q dependence of the shear bias using the FQ method under different assumptions about the galaxy size, noise level, and the PSF form (Gaussian, Moffat, Airy Disk, and CSST g/r/i band). We find that the critical value of Q below which the systematic shear bias would reach more than 1\% is strongly determined by the PSF form and the ratio f between the (post-seeing) galaxy size and the PSF size, not so much by the SNR of the source. In general, the Airy-disk-like PSF performs better than the Gaussian-like PSF, mainly because the Airy-disk type PSF has a more compact power spectrum in Fourier space than the Gaussian function, making it less susceptible to aliasing, as shown in Fig.\ref{fig:gma-als-p}. Overall, for Airy-disk type PSF, sub-percent-level accuracy in shear recovery can be reached for f as low as 1.2 and SNR as low as 15 when Q$\approx 2$ (close to the case of CSST). Interestingly, the tolerance for the pixelation effect can be further extended to even smaller values of Q ($\lesssim 1.7$ for Airy disk/CSST). It can be achieved in the FQ method by simply choosing a large re-convolving Gaussian kernel. The main results are shown in fig.\ref{fig:gau_diag}-\ref{fig:ai-c-e-size-R}.

Our work has demonstrated the robustness of the FQ method in dealing with critically sampled or slightly undersampled galaxy images for accurate shear measurement. In future work, we shall test the whole FQ pipeline using simulated CSST CCDs, and discuss a number of other important issues, such as cosmic-ray detection, PSF reconstruction, etc...

\section*{Acknowledgements}
We thank Xiaokai Chen, Qingyang Li and Hongyu Gao for their useful discussion. This work is supported by the National Key Basic Research and Development Program of China (No.2018YFA0404504), the science research grants from the China Manned Space Project with No.CMS-CSST-2021-A01, and the NSFC grants (11673016, 11621303, 11890691, 12073017, 11903082,61905023, U1931210). The computations in this paper were run on the $\pi$2.0 cluster supported by the Center for High Performance Computing and the Gravity Supercomputer at Shanghai Jiao Tong University. 
 
\bibliography{pixelation_effect}{}

\begin{thebibliography}{}
\expandafter\ifx\csname natexlab\endcsname\relax\def\natexlab#1{#1}\fi
\providecommand{\url}[1]{\href{#1}{#1}}
\providecommand{\dodoi}[1]{doi:~\href{http://doi.org/#1}{\nolinkurl{#1}}}
\providecommand{\doeprint}[1]{\href{http://ascl.net/#1}{\nolinkurl{http://ascl.net/#1}}}
\providecommand{\doarXiv}[1]{\href{https://arxiv.org/abs/#1}{\nolinkurl{https://arxiv.org/abs/#1}}}

\bibitem[{Cao {et~al.}(2018)Cao, Gong, Meng, Xu, Chen, Guo, Li, Liu, Xue, Cao,
  Fu, Zhang, Wang, \& Zhan}]{Cao2018}
Cao, Y., Gong, Y., Meng, X.-M., {et~al.} 2018, Monthly Notices of the Royal
  Astronomical Society, 480, 2178, \dodoi{10.1093/mnras/sty1980}

\bibitem[{Fruchter \& Hook(2002)}]{Fruchter_2002}
Fruchter, A.~S., \& Hook, R.~N. 2002, Publications of the Astronomical Society
  of the Pacific, 114, 144, \dodoi{10.1086/338393}

\bibitem[{{Gong} {et~al.}(2019){Gong}, {Liu}, {Cao}, {Chen}, {Fan}, {Li}, {Li},
  {Li}, {Zhang}, \& {Zhan}}]{Gong2019}
{Gong}, Y., {Liu}, X., {Cao}, Y., {et~al.} 2019, \apj, 883, 203,
  \dodoi{10.3847/1538-4357/ab391e}

\bibitem[{{Heymans} {et~al.}(2006){Heymans}, {Van Waerbeke}, {Bacon}, {Berge},
  {Bernstein}, {Bertin}, {Bridle}, {Brown}, {Clowe}, {Dahle}, {Erben}, {Gray},
  {Hetterscheidt}, {Hoekstra}, {Hudelot}, {Jarvis}, {Kuijken}, {Margoniner},
  {Massey}, {Mellier}, {Nakajima}, {Refregier}, {Rhodes}, {Schrabback}, \&
  {Wittman}}]{Heymans2006}
{Heymans}, C., {Van Waerbeke}, L., {Bacon}, D., {et~al.} 2006, \mnras, 368,
  1323, \dodoi{10.1111/j.1365-2966.2006.10198.x}

\bibitem[{{High} {et~al.}(2007){High}, {Rhodes}, {Massey}, \&
  {Ellis}}]{High2007}
{High}, F.~W., {Rhodes}, J., {Massey}, R., \& {Ellis}, R. 2007, \pasp, 119,
  1295, \dodoi{10.1086/523112}

\bibitem[{{Hirata} \& {Seljak}(2003)}]{HirataSeljak2003}
{Hirata}, C., \& {Seljak}, U. 2003, \mnras, 343, 459,
  \dodoi{10.1046/j.1365-8711.2003.06683.x}

\bibitem[{Hirata \& Seljak(2003)}]{Hirata2003}
Hirata, C.~M., \& Seljak, U. 2003, Monthly Notices of the Royal Astronomical
  Society, 343, 459

\bibitem[{{Hoekstra} {et~al.}(1998){Hoekstra}, {Franx}, {Kuijken}, \&
  {Squires}}]{Hoekstra1998}
{Hoekstra}, H., {Franx}, M., {Kuijken}, K., \& {Squires}, G. 1998, \apj, 504,
  636, \dodoi{10.1086/306102}

\bibitem[{Hoekstra \& Jain(2008)}]{Hoekstra2008}
Hoekstra, H., \& Jain, B. 2008, Annual Review of Nuclear and Particle Science,
  58, 99, \dodoi{10.1146/annurev.nucl.58.110707.171151}

\bibitem[{{Huff} \& {Mandelbaum}(2017)}]{Huff-metacal}
{Huff}, E., \& {Mandelbaum}, R. 2017, arXiv e-prints, arXiv:1702.02600.
\newblock \doarXiv{1702.02600}

\bibitem[{{Kaiser} {et~al.}(1995){Kaiser}, {Squires}, \&
  {Broadhurst}}]{Kaiser1995}
{Kaiser}, N., {Squires}, G., \& {Broadhurst}, T. 1995, \apj, 449, 460,
  \dodoi{10.1086/176071}

\bibitem[{{Kaiser} {et~al.}(2000){Kaiser}, {Wilson}, \& {Luppino}}]{Kaiser2000}
{Kaiser}, N., {Wilson}, G., \& {Luppino}, G.~A. 2000, arXiv e-prints, astro.
\newblock \doarXiv{astro-ph/0003338}

\bibitem[{{Kannawadi} {et~al.}(2021){Kannawadi}, {Rosenberg}, \&
  {Hoekstra}}]{Kannawadi2021}
{Kannawadi}, A., {Rosenberg}, E., \& {Hoekstra}, H. 2021, \mnras, 502, 4048,
  \dodoi{10.1093/mnras/stab211}

\bibitem[{{Kilbinger}(2015)}]{Kilbinger2015}
{Kilbinger}, M. 2015, Reports on Progress in Physics, 78, 086901,
  \dodoi{10.1088/0034-4885/78/8/086901}

\bibitem[{{Laureijs} {et~al.}(2011){Laureijs}, {Amiaux}, {Arduini},
  {Augu{\`e}res}, {Brinchmann}, {Cole}, {Cropper}, {Dabin}, {Duvet}, {Ealet},
  {Garilli}, {Gondoin}, {Guzzo}, {Hoar}, {Hoekstra}, {Holmes}, {Kitching},
  {Maciaszek}, {Mellier}, {Pasian}, {Percival}, {Rhodes}, {Saavedra Criado},
  {Sauvage}, {Scaramella}, {Valenziano}, {Warren}, {Bender}, {Castander},
  {Cimatti}, {Le F{\`e}vre}, {Kurki-Suonio}, {Levi}, {Lilje}, {Meylan},
  {Nichol}, {Pedersen}, {Popa}, {Rebolo Lopez}, {Rix}, {Rottgering},
  {Zeilinger}, {Grupp}, {Hudelot}, {Massey}, {Meneghetti}, {Miller}, {Paltani},
  {Paulin-Henriksson}, {Pires}, {Saxton}, {Schrabback}, {Seidel}, {Walsh},
  {Aghanim}, {Amendola}, {Bartlett}, {Baccigalupi}, {Beaulieu}, {Benabed},
  {Cuby}, {Elbaz}, {Fosalba}, {Gavazzi}, {Helmi}, {Hook}, {Irwin}, {Kneib},
  {Kunz}, {Mannucci}, {Moscardini}, {Tao}, {Teyssier}, {Weller}, {Zamorani},
  {Zapatero Osorio}, {Boulade}, {Foumond}, {Di Giorgio}, {Guttridge}, {James},
  {Kemp}, {Martignac}, {Spencer}, {Walton}, {Bl{\"u}mchen}, {Bonoli},
  {Bortoletto}, {Cerna}, {Corcione}, {Fabron}, {Jahnke}, {Ligori}, {Madrid},
  {Martin}, {Morgante}, {Pamplona}, {Prieto}, {Riva}, {Toledo}, {Trifoglio},
  {Zerbi}, {Abdalla}, {Douspis}, {Grenet}, {Borgani}, {Bouwens}, {Courbin},
  {Delouis}, {Dubath}, {Fontana}, {Frailis}, {Grazian}, {Koppenh{\"o}fer},
  {Mansutti}, {Melchior}, {Mignoli}, {Mohr}, {Neissner}, {Noddle}, {Poncet},
  {Scodeggio}, {Serrano}, {Shane}, {Starck}, {Surace}, {Taylor},
  {Verdoes-Kleijn}, {Vuerli}, {Williams}, {Zacchei}, {Altieri}, {Escudero
  Sanz}, {Kohley}, {Oosterbroek}, {Astier}, {Bacon}, {Bardelli}, {Baugh},
  {Bellagamba}, {Benoist}, {Bianchi}, {Biviano}, {Branchini}, {Carbone},
  {Cardone}, {Clements}, {Colombi}, {Conselice}, {Cresci}, {Deacon}, {Dunlop},
  {Fedeli}, {Fontanot}, {Franzetti}, {Giocoli}, {Garcia-Bellido}, {Gow},
  {Heavens}, {Hewett}, {Heymans}, {Holland}, {Huang}, {Ilbert}, {Joachimi},
  {Jennins}, {Kerins}, {Kiessling}, {Kirk}, {Kotak}, {Krause}, {Lahav}, {van
  Leeuwen}, {Lesgourgues}, {Lombardi}, {Magliocchetti}, {Maguire}, {Majerotto},
  {Maoli}, {Marulli}, {Maurogordato}, {McCracken}, {McLure}, {Melchiorri},
  {Merson}, {Moresco}, {Nonino}, {Norberg}, {Peacock}, {Pello}, {Penny},
  {Pettorino}, {Di Porto}, {Pozzetti}, {Quercellini}, {Radovich}, {Rassat},
  {Roche}, {Ronayette}, {Rossetti}, {Sartoris}, {Schneider}, {Semboloni},
  {Serjeant}, {Simpson}, {Skordis}, {Smadja}, {Smartt}, {Spano}, {Spiro},
  {Sullivan}, {Tilquin}, {Trotta}, {Verde}, {Wang}, {Williger}, {Zhao},
  {Zoubian}, \& {Zucca}}]{Laureijs2011}
{Laureijs}, R., {Amiaux}, J., {Arduini}, S., {et~al.} 2011, arXiv e-prints,
  arXiv:1110.3193.
\newblock \doarXiv{1110.3193}

\bibitem[{{Mandelbaum}(2018)}]{Mandelbaum2018}
{Mandelbaum}, R. 2018, \araa, 56, 393,
  \dodoi{10.1146/annurev-astro-081817-051928}

\bibitem[{{Mandelbaum} {et~al.}(2006){Mandelbaum}, {Hirata}, {Ishak}, {Seljak},
  \& {Brinkmann}}]{Mandelbaum2006a}
{Mandelbaum}, R., {Hirata}, C.~M., {Ishak}, M., {Seljak}, U., \& {Brinkmann},
  J. 2006, \mnras, 367, 611, \dodoi{10.1111/j.1365-2966.2005.09946.x}

\bibitem[{Rowe {et~al.}(2011)Rowe, Hirata, \& Rhodes}]{Rowe_2011}
Rowe, B., Hirata, C., \& Rhodes, J. 2011, The Astrophysical Journal, 741, 46,
  \dodoi{10.1088/0004-637x/741/1/46}

\bibitem[{{Rowe} {et~al.}(2015){Rowe}, {Jarvis}, {Mandelbaum}, {Bernstein},
  {Bosch}, {Simet}, {Meyers}, {Kacprzak}, {Nakajima}, {Zuntz}, {Miyatake},
  {Dietrich}, {Armstrong}, {Melchior}, \& {Gill}}]{Rowe2015}
{Rowe}, B.~T.~P., {Jarvis}, M., {Mandelbaum}, R., {et~al.} 2015, Astronomy and
  Computing, 10, 121, \dodoi{10.1016/j.ascom.2015.02.002}

\bibitem[{Shapiro {et~al.}(2013)Shapiro, Rowe, Goodsall, Hirata, Fucik, Rhodes,
  Seshadri, \& Smith}]{Shapiro_2013}
Shapiro, C., Rowe, B. T.~P., Goodsall, T., {et~al.} 2013, Publications of the
  Astronomical Society of the Pacific, 125, 1496, \dodoi{10.1086/674415}

\bibitem[{{Sheldon} \& {Huff}(2017)}]{Sheldon-metacal}
{Sheldon}, E.~S., \& {Huff}, E.~M. 2017, \apj, 841, 24,
  \dodoi{10.3847/1538-4357/aa704b}

\bibitem[{Wang {et~al.}(2021)Wang, Zhang, Li, \& Shen}]{Wang_2021}
Wang, H., Zhang, J., Li, H., \& Shen, Z. 2021, The Astrophysical Journal, 911,
  10, \dodoi{10.3847/1538-4357/abe856}

\bibitem[{{Zellem} {et~al.}(2022){Zellem}, {Nemati}, {Bailey}, {Cady},
  {Colavita}, {Gonzalez}, {Hildebrandt}, {Maier}, {Mennesson}, {Ygouf},
  {Zimmerman}, {Belikov}, {Debes}, {De Rosa}, {Douglas}, {Girard}, {Groff},
  {Kasdin}, {Lowrance}, {Macintosh}, {Payne}, {Ryan}, \& {Weisberg}}]{roman}
{Zellem}, R.~T., {Nemati}, B., {Bailey}, V.~P., {et~al.} 2022, arXiv e-prints,
  arXiv:2202.05923.
\newblock \doarXiv{2202.05923}

\bibitem[{{Zhan}(2011)}]{Zhan2011}
{Zhan}, H. 2011, Scientia Sinica Physica, Mechanica \& Astronomica, 41, 1441,
  \dodoi{10.1360/132011-961}

\bibitem[{{Zhang}(2008)}]{Zhang2008}
{Zhang}, J. 2008, \mnras, 383, 113, \dodoi{10.1111/j.1365-2966.2007.12585.x}

\bibitem[{{Zhang}(2010)}]{Zhang2010}
---. 2010, \mnras, 403, 673, \dodoi{10.1111/j.1365-2966.2009.16168.x}

\bibitem[{{Zhang}(2011)}]{zhang2011JCAP}
---. 2011, \jcap, 2011, 041, \dodoi{10.1088/1475-7516/2011/11/041}

\bibitem[{Zhang \& Komatsu(2011)}]{Zhang2011}
Zhang, J., \& Komatsu, E. 2011, Monthly Notices of the Royal Astronomical
  Society, 414, 1047, \dodoi{10.1111/j.1365-2966.2011.18436.x}

\bibitem[{{Zhang} {et~al.}(2015){Zhang}, {Luo}, \& {Foucaud}}]{Zhang2015}
{Zhang}, J., {Luo}, W., \& {Foucaud}, S. 2015, \jcap, 2015, 024,
  \dodoi{10.1088/1475-7516/2015/01/024}

\bibitem[{Zhang {et~al.}(2016)Zhang, Zhang, \& Luo}]{Zhang_2016}
Zhang, J., Zhang, P., \& Luo, W. 2016, The Astrophysical Journal, 834, 8,
  \dodoi{10.3847/1538-4357/834/1/8}

\bibitem[{{Zhang} {et~al.}(2017){Zhang}, {Zhang}, \& {Luo}}]{Zhang2017PDF}
{Zhang}, J., {Zhang}, P., \& {Luo}, W. 2017, \apj, 834, 8,
  \dodoi{10.3847/1538-4357/834/1/8}

\bibitem[{Zhang {et~al.}(2019)Zhang, Dong, Li, Li, Li, Liu, Luo, Fu, Li, \&
  Fan}]{Zhang_2019}
Zhang, J., Dong, F., Li, H., {et~al.} 2019, The Astrophysical Journal, 875, 48,
  \dodoi{10.3847/1538-4357/ab1080}

\end{thebibliography}
\bibliographystyle{aasjournal}

\end{document}